\documentclass[a4paper,fleqn,usenatbib]{mnras}

\usepackage{graphicx}	
\usepackage{amsmath}	
\usepackage{amssymb}	
\usepackage[capitalise]{cleveref}
     \crefname{equation}{Equation}{Equations}
     \crefname{figure}{Figure}{Figures}
     \crefname{table}{Table}{Tables}
     
\usepackage{enumitem}
\usepackage{calc}
\usepackage{siunitx}
\usepackage{booktabs}
\usepackage{flushend}
\usepackage[T1]{fontenc}
\usepackage{array}
\usepackage{makecell}
\usepackage{tikz}
\usepackage{movie15}
\usepackage{animate}

\allowdisplaybreaks

\DeclareFontFamily{OT1}{pzc}{}
\DeclareFontShape{OT1}{pzc}{m}{it}{<-> s * [1.10] pzcmi7t}{}
\DeclareMathAlphabet{\mathpzc}{OT1}{pzc}{m}{it}

\usepackage[T1]{fontenc}
\usepackage{ae,aecompl}

\newcommand\diag[4]{%
  \multicolumn{1}{p{#2} }{\hskip-\tabcolsep 
  $\vcenter{\begin{tikzpicture}[baseline=0,anchor=south west,inner sep=#1]
  \path[use as bounding box] (0,0) rectangle (#2+2\tabcolsep,\baselineskip);
  \node[minimum width={#2+2\tabcolsep-\pgflinewidth},
        minimum  height=\baselineskip+\extrarowheight-\pgflinewidth] (box) {};
  \draw[line cap=round] (box.north west) -- (box.south east);
  \node[anchor=south west] at (box.south west) {#3};
  \node[anchor=north east] at (box.north east) {#4};
 \end{tikzpicture}}$\hskip-\tabcolsep}}

\newcolumntype{x}[1]{>{\centering\arraybackslash}p{#1}}

\usepackage{natbib}
\defcitealias{Laibe/Price/2014a}{LP14a}
\defcitealias{Laibe/Price/2014b}{LP14b}

\defcitealias{Hutchison/etal/2016}{HPLM16}

\title[On the maximum entrainable grain size]{On the maximum grain size entrained by photoevaporative winds}

\author[Hutchison, Laibe, \& Maddison]{
Mark A. Hutchison,$^{1}$\thanks{E-mail: mhutchison@swin.edu.au}
Guillaume Laibe,$^{2}$
and Sarah T. Maddison$^{1}$
\\
$^{1}$Centre for Astrophysics \& Supercomputing, Swinburne University of Technology, Hawthorn, VIC 3122, Australia\\
$^{2}$School of Physics and Astronomy, University of St. Andrews, North Haugh, St. Andrews, Fife KY16 9SS, UK
}

\date{Accepted XXX. Received YYY; in original form ZZZ}

\pubyear{2016}

\begin{document}
\label{firstpage}
\pagerange{\pageref{firstpage}--\pageref{lastpage}}
\maketitle

\begin{abstract}
We model the behaviour of dust grains entrained by photoevaporation-driven winds from protoplanetary discs assuming a non-rotating, plane-parallel disc. We obtain an analytic expression for the maximum entrainable grain size in extreme-UV radiation-driven winds, which we demonstrate to be proportional to the mass loss rate of the disc. When compared with our hydrodynamic simulations, the model reproduces almost all of the wind properties for the gas and dust. In typical turbulent discs, the entrained grain sizes in the wind are smaller than the theoretical maximum everywhere but the inner disc due to dust settling.
\end{abstract}

\begin{keywords}
planets and satellites: atmospheres -- protoplanetary discs -- circumstellar matter -- stars: pre-main-sequence
\end{keywords}



\section{Introduction}
\label{sec:introduction}

Small dust grains in the upper atmosphere dominate the opacity for the disc at many wavelengths, thereby shielding the bulk of the disc from energetic radiation from the star and controlling the disc's thermal/geometric structure \citep{Calvet/etal/1991,Chiang/Goldreich/1997,DAlessio/etal/1998}. Imaging scattered starlight and thermal re-emission of absorbed stellar radiation from dust in these upper layers is still a vital diagnostic tool used to characterise discs and their structure \citep{Watson/etal/2007,Andrews/2015}. It therefore follows that physical processes that affect the dynamics of these grains (e.g. settling, grain growth, and disc winds) may have an impact in the way that we interpret observations of protoplanetary discs \citep{testi/etal/2014}.

Aerodynamic drag from disc winds can loft dust into the atmospheres surrounding discs. Using order-of-magnitude force balance arguments, \citet{Takeuchi/Clarke/Lin/2005} estimate the maximum grain size that can be carried out by photoevaporative winds. Better estimates were obtained by \citet{Owen/Ercolano/Clarke/2011a} who test \emph{a posteriori} whether dust can be entrained along gas streamlines in single-phase photoevaporation simulations. More recently, we have performed fully coupled, gas and dust hydrodynamic simulations of protoplanetary discs undergoing dust settling and extreme ultraviolet (EUV) induced photoevaporation \citep[][hereafter, \citetalias{Hutchison/etal/2016}]{Hutchison/etal/2016}. Based on the suite of simulations for that study, we concluded that only micron sized dust grains and smaller are entrained by photoevaporative winds in typical discs found around T Tauri stars. The exact cutoff, however, was found to depend on stellar mass, stellar irradiation flux, gas density at the base of the flow, and distance from the central star.

Numerical simulations of gas and dust potentially provide one of the best windows on dust dynamics in discs, but owing to the numerical difficulty associated with simulating small dust grains in such steeply stratified atmospheres, they are still too unwieldy to use in a practical sense for global disc studies across multiple systems. A nice alternative to using numerical simulations is the self-similar solution for thermal disc winds derived by \citet{Clarke/Alexander/2016}. In this study, we provide another alternative by deriving an easy to use (semi-)analytic solution that recovers the majority of the results from our hydrodynamic simulations.

The paper is organised as follows: in \cref{sec:semi-analytic_model} we derive the equations for our semi-analytic model and compare the model with hydrodynamic simulations; in \cref{sec:results} we use our model to explore different parameters that affect dust entrainment in disc winds; in \cref{sec:effects_of_settling} we discuss the effects of settling; and in \cref{sec:conclusions} we summarise our findings.


\section{Semi-analytic dusty wind model}
\label{sec:semi-analytic_model}

Previously, \citet{Hutchison/Laibe/2016} derived an analytic solution for EUV-driven winds assuming a non-rotating, plane-parallel atmosphere. The simple geometry makes the problem tractable, retains the vertical disc structure, and reproduces the vertical winds near the ionisation front. Later, \citetalias{Hutchison/etal/2016} showed using hydrodynamic simulations that the back reaction on outflowing gas due to entrained dust grains is negligible due to the small dust-to-gas ratios in the upper atmospheres of discs. We exploit this fact to extend our model to two fluids by directly inserting the analytic wind solution for the gas into the fluid equations for dust.

\subsection{Gas}
\label{sec:gas}

We assume an isothermal thin disc supported by pressure-gravity balance due to the vertical component of gravity from a central star
\begin{equation}
	\mathbf{g} = -\frac{\mathcal{G} M z}{\left( R^2+z^2 \right)^{3/2}}\mathbf{\hat{z}},
	\label{eq:vertical_gravity}
\end{equation}
where $\mathcal{G}$ is the gravitational constant, $M$ is the mass of the central star, and $z$ is the height above the midplane. In this geometry, the parameter $R$ and the variable $z$ make up a quasi-2D coordinate system centred on the star. The wind speed for isothermal photoevaporation can be written in closed form using the Lambert $\mathrm{W}$ function \citep{Corless/etal/1996,Veberic/2012},
\begin{equation}
	v_\text{g} = c_\text{s} \sqrt{- \, \mathrm{W}_0 \! \! \left[ -\exp{ \left( -\frac{2\mathcal{G}M}{c_\text{s}^2 \sqrt{R^2+z^2}} - 1 \right)  }   \right]},
	\label{eq:plane-parallel_solution}
\end{equation}
where $v_\text{g}$ is the gas velocity and $c_\text{s}\approx 10\,\text{km}\,\text{s}^{-1}$ is the isothermal sound speed of the wind. The gas density is related to the velocity via the relation
\begin{equation}
	\dot{m}_\text{g} = \rho_\text{g} v_\text{g},
	\label{eq:accretion_rate}
\end{equation}
where $\dot{m}$ is the constant mass loss rate per unit area of the outflow in a stationary regime.

The value of $\dot{m}$ is best determined by the fluid quantities at the base of the flow. The gas velocity in the wind is well constrained by \cref{eq:plane-parallel_solution}, so this amounts to determining $\rho_\text{g,i}$. For simplicity, we will assume the density in the disc and wind is piecewise continuous, but the reality is that collisional heating from the ionised wind will distort the density structure of the neutral disc near the ionisation front, causing the density to decrease faster than if photoevaporation was not present. We have not performed an exhaustive study of how $\rho_\text{g,i}$ changes with stellar and disc parameters, but the suite of simulations performed by \citetalias{Hutchison/etal/2016} show that collisional heating from the wind causes the initial outflow to level off at $\sim 40\%$ of the assumed ionisation front density. Neglecting this density offset equates to overestimating dust entrainment in the wind (see \cref{sec:base_flow_density}). However, \citetalias{Hutchison/etal/2016} also showed that a non-rotating, plane-parallel atmosphere underestimates dust entrainment by approximately the same amount. As a result, we only worry about scaling the density when comparing our model directly with numerical simulations.

\subsection{Dust}
\label{sec:dust}

\subsubsection{Equations of motion}
\label{sec:equations_of_motion}

In the limit of small dust-to-gas ratio, the steady-state fluid equations for the dust are simplified to the following equations:
\begin{align}
	& \nabla \cdot (\rho_\text{d} \mathbf{v}_\text{d})  = 0,
	\label{eq:dust_continuity}
\\
	& \mathbf{v}_\text{d} \cdot \nabla \mathbf{v}_\text{d}  = \frac{1}{t_\text{s}} \left( \mathbf{v}_\text{g}-\mathbf{v}_\text{d} \right) +  \mathbf{g},
	\label{eq:dust_momentum}
\end{align}
where $t_\text{s}$ is the Epstein drag stopping time \citep{Epstein/1924},
\begin{equation}
	t_\text{s} =  \sqrt{\frac{\pi \gamma}{8}} \frac{\rho_\text{grain} s }{c_\text{s} \rho_\text{g}} = \frac{\rho_\text{eff} s }{c_\text{s} \rho_\text{g}},
	\label{eq:epstein drag}
\end{equation}
and $s$ and $\rho_\text{grain}$ are respectively the intrinsic size and density of the individual dust grains. For convenience, we simplify the expression in the second equality by defining $\rho_\text{eff} \equiv \rho_\text{grain} \sqrt{\pi \gamma/8}$ as an effective grain density.

The continuity equation \labelcref{eq:dust_continuity} is identical in form to that of the gas so the solution can be read directly from \cref{eq:accretion_rate},
\begin{equation}
	\dot{m}_\text{d} = \rho_\text{d} v_\text{d}.
	\label{eq:dust_accretion_rate}
\end{equation}
Meanwhile, the momentum equation \labelcref{eq:dust_momentum} reduces to a single ordinary differential equation
\begin{equation}
	v_\text{d} \frac{\text{d} v_\text{d}}{\text{d}z} = \frac{1}{t_\text{s}} \left( v_\text{g} - v_\text{d} \right) - \frac{\mathcal{G} M z}{\left( R^2 + z^2 \right)^{3/2}},
	\label{eq:dust_ode_with_dim}
\end{equation}
where $v_\text{g}$ is a function of $z$ and is given by \cref{eq:plane-parallel_solution}. \Cref{eq:dust_ode_with_dim} can be written in dimensionless form using the parameters $\bar{v} \equiv v/v_\text{K}$, $\bar{z} \equiv z/R$, and $v_\text{K} = \sqrt{\mathcal{G} M/R}$:   
\begin{equation}
	\bar{v}_\text{d} \frac{\text{d} \bar{v}_\text{d}}{\text{d} \bar{z}} = \mathcal{S}\!\mathpzc{t}^{-1} \left( 1 - \frac{\bar{v}_\text{d}}{\bar{v}_\text{g}} \right) - \frac{\bar{z}}{\left( 1 + \bar{z}^2 \right)^{3/2}},
	\label{eq:dust_ode}
\end{equation}
where, by analogy to dust dynamics in a disc, we have defined the Stokes number of the wind to be
\begin{equation}
	\mathcal{S}\!\mathpzc{t} \equiv \frac{\rho_\text{eff} s v_\text{K}^{2}}{c_\text{s} \dot{m}_\text{g} R} = \frac{v_\text{K}^{2}/R}{v_\text{g,i} / t_\text{s,i}},
	\label{eq:stokes_num}
\end{equation}
where $v_\text{g,i} $, $\rho_\text{g,i} $, and $ t_\text{s,i} = \rho_\text{eff} s / c_\text{s} \rho_\text{g,i}$ denote the gas velocity, the gas density and the stopping time at the base of the flow, respectively. We emphasise that the Stokes number in the wind, denoted here as $\mathcal{S}\!\mathpzc{t}$, is \emph{a priori} different to the Stokes number in the disc ($\text{St}$). \Cref{eq:stokes_num} can be seen as the ratio between the gravitational force and the force required to keep the grains entrained at the base of the flow.

\subsubsection{Asymptotic behaviour}
\label{sec:asymptotic_behaviour}

When $\bar{z} \to \infty$, the term $-\bar{z} / (1 + \bar{z}^{2})^{3/2} \to 0$ while \cref{eq:plane-parallel_solution} implies that $\bar{v}_\text{g} \to 1$ and $\text{d} \bar{v}_\text{g}/\text{d} \bar{z} \to 0$. Upon applying these limits to \cref{eq:dust_ode}, the solution $\bar{v}_\text{d} \simeq \bar{v}_\text{g}$ with a vanishing derivative can readily by seen by inspection. Note that this holds for all Stokes numbers. On the other hand, when $\bar{z} \to \bar{z}_\text{i}$, where $\bar{z}_\text{i}$ is the location of the initial flow, the Stokes number determines the behaviour of the solution. Possible solutions can be categorised into two main classes based on whether the dust velocity is initially increasing or decreasing.
\begin{description}
\item[Increasing:]
For $\mathcal{S}\!\mathpzc{t} \to 0$ (i.e. high drag), the positive drag term dominates over the negative gravitational component, thus implying that $\text{d} \bar{v}_\text{d} / \text{d} \bar{z}  > 0$ for all $\bar{z}$. However, to keep the drag term from becoming unbounded, the gas and dust velocities must be approximately equal, to zeroth order in $\mathcal{S}\!\mathpzc{t}$. Two distinct subclasses of grains are possible: 
\begin{enumerate}[leftmargin=1.75\parindent]
	\item \emph{perfectly}-entrained grains that adhere to the zeroth order approximation, and
	\item \emph{well}-entrained grains that do not.
\end{enumerate}
\item[Decreasing:]
For $\mathcal{S}\!\mathpzc{t} \to \infty$ (i.e. low drag), the solution satisfies, to zeroth order in $\mathcal{S}\!\mathpzc{t}^{-1}$,
\begin{equation}
	\bar{v}_\text{d} \frac{\partial \bar{v}_\text{d}}{\partial \bar{z}} \simeq -  \frac{\bar{z}}{\left( 1 + \bar{z}^{2} \right)^{3/2}}.
\end{equation}
Fully entrained flows require $\bar{v}_\text{d} > 0$, but allowing $\bar{v}_\text{d}$ and $\text{d} \bar{v}_\text{d}/\text{d} \bar{z}$ to be general yields a total of three new subclasses:
\begin{enumerate}[leftmargin=1.75\parindent]
	\setcounter{enumi}{2}
	\item \emph{weakly}-entrained dust grains with positive velocities throughout the flow (i.e. away from the midplane),
	\item \emph{partially}-entrained dust grains whose velocities change sign in the flow, and
	\item \emph{non}-entrained dust grains whose initial velocities are negative (i.e. toward the midplane).
\end{enumerate}
Note that these latter two subclasses are best interpreted with an Eulerian perspective, since at large $\bar{z}$ the velocities will always become positive and converge to the gas velocity. Although a steady state flow is impossible to achieve if $\bar{v}_\text{d}$ reverses direction, this would imply a pile-up of some kind for partially-entrained dust grains. Moreover, the distinction between weakly- and partially-entrained dust grains suggests there exists a critical Stokes number $\mathcal{S}\!\mathpzc{t}_\text{c}$ for which the minimum of $\bar{v}_\text{d}$ is zero. This ensures that when $\mathcal{S}\!\mathpzc{t} < \mathcal{S}\!\mathpzc{t}_\text{c}$ particles are entrained by the photoevaporative wind.
\end{description}

\subsubsection{Maximum entrained grain size}
\label{sec:critical_seize}

We denote $\bar{z}_\text{c}$ as the height at which $\bar{v}_\text{d} = \text{d} \bar{v}_\text{d} / \text{d} \bar{z} = 0$. Substituting these values into \cref{eq:dust_ode} yields the following relation,
\begin{equation}
	\frac{\mathcal{S}\!\mathpzc{t}_\text{c} \bar{z}_\text{c}}{\left(1 + \bar{z}_\text{c}^{2} \right)^{3/2}} = 1.
	\label{eq:crit_condition}
\end{equation}
Taking the derivative of \cref{eq:crit_condition} with respect to $\bar{z}_\text{c}$ removes the $\mathcal{S}\!\mathpzc{t}_\text{c}$ dependence and allows us to solve for $\bar{z}_\text{c}$,
\begin{equation}
	\bar{z}_\text{c} = \pm \frac{1}{\sqrt{2}}. 
	\label{eq:zcrit}
\end{equation}
This is also the location for the peak gravitational force. Substituting $\bar{z}_\text{c}$ back into \cref{eq:crit_condition} gives us the critical Stokes number,
\begin{equation}
	\mathcal{S}\!\mathpzc{t}_\text{c} = \frac{3 \sqrt{3}}{2} \simeq 2.6.
\end{equation}
From \cref{eq:stokes_num}, we can then solve for the maximum entrainable grain size in the wind,
\begin{equation}
	s_\text{max} = \frac{3 \sqrt{3}}{2} \frac{c_\text{s} \dot{m}_\text{g} R}{\rho_\text{eff} v_\text{K}^{2}} ,
	\label{eq:smax}
\end{equation}
where $\dot{m}_\text{g}$ is obtained from \cref{eq:accretion_rate} using initial conditions at the base of the flow. The critical height $\bar{z}_\text{c}$ is close enough to the disc surface for the plane-parallel approximation to remain valid, but is subject to the validity of the assumptions made about the underlying disc.

We can alternatively derive this limit from a Lagrangian perspective by retaining the $\text{d}v_\text{d}/\text{d}t$ term in \cref{eq:dust_ode_with_dim} and rewriting the equation using an effective potential,
\begin{equation}
	V_\text{eff} \equiv -\left( \frac{\bar{z}}{\mathcal{S}\!\mathpzc{t}}  + \frac{1}{\sqrt{1+\bar{z}^2}} \right),
	\label{eq:effective_potential}
\end{equation}
such that
\begin{equation}
	\frac{\text{D}^2\bar{z}}{\text{D}\bar{t}^2} = -\frac{1}{\mathcal{S}\!\mathpzc{t} \bar{v}_\text{g}} \frac{\text{D}\bar{z}}{\text{D}\bar{t}} - \frac{\text{d}V_\text{eff} }{\text{d}\bar{z}},
	\label{eq:lagrangian_dust_ode}
\end{equation}
where the convective derivative ($\text{D/D}\bar{t} \equiv \partial/\partial \bar{t} + \bf{\bar{v}} \cdot \nabla$) is taken with respect to the dimensionless time variable, $\bar{t} \equiv \Omega_\text{K} t$, with $\Omega_\text{K} = v_\text{K}/R$.  Except when $\mathcal{S}\!\mathpzc{t}>\mathcal{S}\!\mathpzc{t}_\text{c}$, the new effective potential monotonically decreases to infinity (i.e. dust grains are entrained and escape the system). Above $\mathcal{S}\!\mathpzc{t}_\text{c}$ a local minimum forms that keeps grains with $s > s_\text{max}$ bound to the disc. The important point here is that $V_\text{eff}$ is independent of the functional form of $v_\text{g}$ as long as $v_\text{g} \propto 1/\rho_\text{g}$, or equivalently, as long as the continuity equation is valid. 

\subsubsection{Underlying disc structure}
\label{sec:underlying_disc}

We must integrate \cref{eq:dust_ode} numerically to find the dust velocity, but this requires initial conditions for the flow, including the gas and dust structure in the underlying disc. \Cref{eq:plane-parallel_solution} specifies the gas velocity for all $z$, but does not identify where the flow begins. Physically, the ionisation front location marking the base of the flow is set by the intensity of the impinging radiation field and its corresponding optical depth. For maximum flexibility, we leave $\rho_\text{g,i}$ as an input parameter for our model and parameterise the penetration depth of the EUV radiation by defining $\xi \equiv \rho_\text{g,i}/ \rho_\text{g,0}$, where $\rho_\text{g,0}$ is the local midplane density of the disc. The location of the ionisation front is then obtained by solving for the height $z_\text{i}$ at which the disc density is equal to $\rho_\text{g,i}$.
 
For the density profile of the disc we use the isothermal thin disc approximation
\begin{equation}
	\rho_\text{g}(z) = \rho_{\text{g},0} \exp{\left[ -\frac{ z^2}{2 H^2}\right]},
	\label{eq:disc_density}
\end{equation}
where $H$ is the local scale height of the disc. We specify the entire $R$-$z$ disc structure using a power-law parameterisation \citep[see, e.g.,][]{Laibe/Gonzalez/Maddison/2012}
\begin{align}
	& \Sigma_\text{g} = \Sigma_{\text{g},1\text{AU}} \left( \frac{R}{1\,\text{AU}} \right)^{-p},
\\
	& H = H_{1\text{AU}} \left( \frac{R}{1\,\text{AU}} \right)^{3/2- q/2},
\\
	&\rho_{\text{g,0}} =  \frac{\Sigma_\text{g}}{\sqrt{2\pi}H},
\end{align}
where $\Sigma_\text{g}$ is the local surface density for the gas while quantities with the subscript $1\,$AU are reference values measured at $R=1\,$AU. The parameters $p$ and $q$ are power-law exponents controlling the density and temperature (i.e. flaring) of the disc, respectively. Observations and simulations indicate that $p$ and $q$ can cover a range of values -- $q$ being the better constrained out of the two \citep[e.g.][]{Dutrey/etal/1996,Andrews/Williams/2005,Andrews/Williams/2007,Laibe/Gonzalez/Maddison/2012,Pinte/Laibe/2014}. In keeping with these studies, we adopt the ranges $p \in [0,1.5]$ and $q \in [0.4,0.8]$.
 
We assume all grain sizes are uniformly distributed throughout the disc with the same density structure as the gas, only scaled by the dust-to-gas ratio $\varepsilon = 0.01$. While unphysical, distributing the dust in this manner makes it much easier to compare different grain sizes, identify trends, and gain valuable insight about how dust behaves in discs. The only caveat is that we partially lose the ability to make predictions about dust in real outflows, particularly for larger grains which we would expect to be concentrated closer to the midplane and away from the ionisation front. In fact, from \citetalias{Hutchison/etal/2016} we know that settling can keep dust grains that are at least a few times smaller than $s_\text{max}$ from ever entering the wind. Correctly accounting for this behaviour requires hydrodynamic simulations or assuming a steady-state stratified disc structure for the dust (see \cref{sec:effects_of_settling}).

One may intuitively expect the initial dust velocity to start from rest, but this is physically inconsistent with the relation $\rho_\text{d} = \dot{m}_\text{d}/v_\text{d}$, which would create an unphysical density spike at the surface of the disc. The numerical models from \citetalias{Hutchison/etal/2016} show that the ionisation front is a very narrow, dynamically complicated transition region where neither phase is particularly well represented by the solutions above. Once the flow has settled, the dust already has a finite velocity and the solution is better represented using a zero derivative at $\bar{z}=\bar{z}_\text{i}$. Substituting this into \cref{eq:dust_ode} and solving for $\bar{v}_\text{d}$ results in the following initial condition for the dust
\begin{equation}
	\bar{v}_\text{d,i} = \bar{v}_\text{g,i} \left[ 1 - \frac{\mathcal{S}\!\mathpzc{t} \, \bar{z}}{\left(1 + \bar{z}^2 \right)^{3/2}} \right].
\end{equation}

\subsection{Verifying $s_\text{max}$}
\label{sec:verifying_smax}

We verify that $s_\text{max}$ is valid by comparing the semi-analytic model here with the hydrodynamic model in \citetalias{Hutchison/etal/2016}. In doing so, we adopt the following values for our disc: $R = 5\,$AU, $M = 1\,M_\odot$, $\Sigma_{\text{g},1\text{AU}} = 100\,\text{g}\,\text{cm}^{-2}$, $H_{1\text{AU}} = 0.05\,$AU, $p = 1$, and $q = 0.5$. We assume the EUV penetration depth is $\xi = 10^{-5}$, which is consistent with thermo-chemical models of typical T Tauri discs \citep{Woitke/etal/2016}. In the ionised wind, gas particles are held isothermally (i.e. $\gamma  = 1$) at $T = 10^4\,$K such that $c_\text{s} \approx 10\,\text{km}\,\text{s}^{-1}$. Finally, we assume that the intrinsic dust density for all our dust grains is $\rho_\text{grain} = 3\,\text{g}\,\text{cm}^{-3}$. Plugging these values into \cref{eq:smax} give $s_\text{max} \simeq 0.82\,\mu$m. Except where noted otherwise, we will use the fiducial values above for the remainder of the paper. 

Although the outflow in this geometry is 1D in nature, the transition between disc and outflow is better captured in multi-dimensional simulations. This is because ionisation of particles at the disc surface causes compression of adjacent neutral particles, which in 1D produces sporadic bursts of outflow. Thus for our numerical simulation, we place $200\,028$ particles on a uniform (staggered) lattice inside a Cartesian box, $(x,z) \in [-1.9 , 1.9]\,$AU and set the gas/dust masses and dust fraction using the method described in \citetalias{Hutchison/etal/2016} for unequal-mass, one-fluid particles. We use periodic boundary conditions in $x$ and dynamic boundaries in $z$, consistent with a steady-state wind flowing away from the disc \citep{Hutchison/Laibe/2016}. We create the dynamic boundaries by converting the ionised particles at $t=0$ into ghost particles and forcing them to move in the vertical direction at the local wind speed prescribed by \cref{eq:plane-parallel_solution}. The number of ghost particles produced by this setup is $5372$ on each side of the disc. Photoevaporation is created by heating gas particles to $10^4\,$K when $\rho_\text{g} \leq \rho_\text{g,i} = \xi \rho_\text{g,0}$. The disc begins the simulation in isothermal hydrostatic equilibrium, but is evolved adiabatically from $t = 0$ to capture the collisional heating at the ionisation front.

\begin{figure}
\includemovie[
  poster=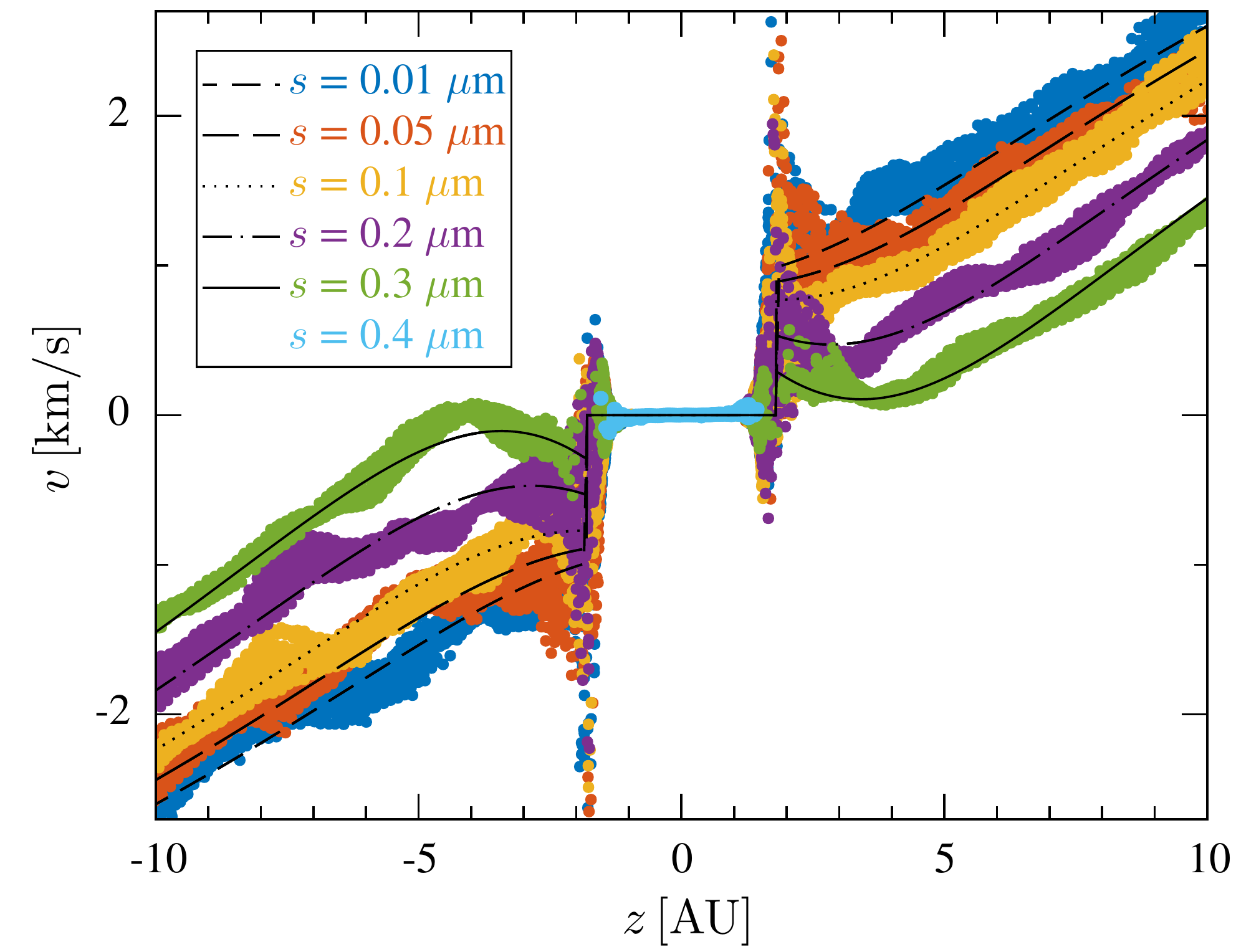,
  text=
]{\columnwidth}{0.75\columnwidth}{compare_velocities.mp4}
	\caption{Dust velocities as a function of $z$ for the 6 indicated grain sizes. The coloured points are results from hydrodynamic simulations at $t = 80\,$yr while the black lines are the corresponding semi-analytic results. The simulations confirm that $s_\text{max}$ is between $0.3$--$0.4\,\mu$m, the predicted value being $0.33\,\mu$m. Note transient oscillations appear in the numerical solution, but always oscillate closely about the semi-analytic solution. (An animated version is playable in the online pdf.)}
	\label{fig:compare_velocities}
\end{figure}

We measure the reduction in $\rho_\text{g,i}$ caused by ionisation front heating in the simulation to be $\sim40\%$, thereby reducing $s_\text{max}$ from $0.82$ to $0.33\,\mu$m. Note, we need not worry about specifying $\rho_\text{d,i}$ because it has no influence in determining the value of $s_\text{max}$. \Cref{fig:compare_velocities} shows a snapshot of the hydrodynamic velocities at $t = 80\,$yr overlaid with the semi-analytic curves assuming a $40\%$ reduction in gas density in the flow. The hydrodynamic solutions are somewhat noisy due to fluctuating motions in the flow produced by the stochastic ionisation of gas particles at the ionisation front, but they always oscillate about their respective semi-analytic solution. The trend of $v_\text{g} \to 0$ as $s \to s_\text{max}$ is a clear indicator that \cref{eq:smax} is correct. As further proof, when we try $s=4\,\mu$m, no dust is entrained in the outflow.


We emphasise that $s_\text{max}$ is a robust, physical limit set by gas properties in the wind and is not affected by any properties of the disc. In fact, the above simulations were run with dust settling enabled to show that dust density in the disc has no effect on the entrainment properties in the wind (unless of course the density is zero, see \cref{sec:effects_of_settling}). Because it does not matter what physical mechanisms supply the dust to the disc surface (e.g. turbulence, migration, accretion jets), $s_\text{max}$ is model independent. This is important because any observational constraint on $s$ in the wind can be converted into a strict lower bound on $\dot{m}_\text{g}$ by inverting \cref{eq:smax}. In terms of surface density, this translates into a photoevaporative mass loss rate of
\begin{equation}
	\dot{\Sigma}_\text{g,photo} \geq \frac{8 \pi}{3 \sqrt{3}} \frac{\rho_\text{eff} s_\text{obs} v_\text{K}^{2}}{c_\text{s}},
	\label{eq:surface_density_mass_loss}
\end{equation}
where $s_\text{obs}$ is the largest grain size observed in the flow.


\section{Dependance on disc parameters}
\label{sec:results}

In this section, we use \cref{eq:smax,eq:dust_ode} to investigate how $s_\text{max}$ and dust entrainment depend on the model's disc and stellar parameters by systematically varying the grain size, disc radius, base flow density, and stellar mass.


\subsection{Grain size}
\label{sec:grain_size}

\begin{figure}
	\centering{\includegraphics[width=\columnwidth]{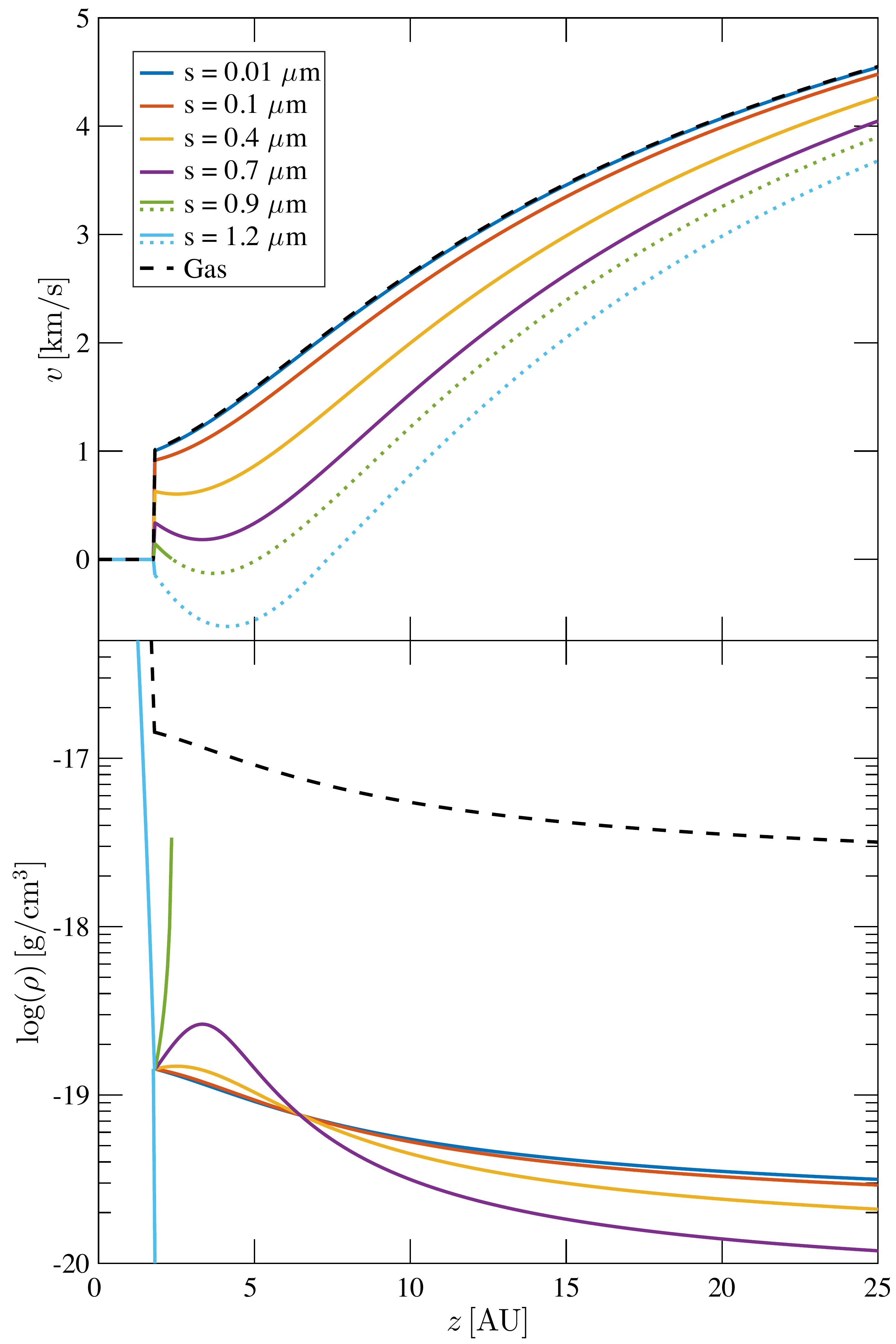}}
	\caption{Velocity (top) and density (bottom) profiles for gas (dashed) and dust (solid) in a photoevaporative wind at $R=5\,$AU. Grains with $s \gtrsim 0.8\,\mu$m cannot be entrained in the wind, but may lead to a pile-up near the surface of the disc. Large velocity differences between phases is an indication of the lack of dust entrainment. The velocity-density relation in \cref{eq:dust_accretion_rate} leads to a seesaw pattern in the density.}
	\label{fig:grain_size}
\end{figure}

Using the values above, we solve \cref{eq:dust_ode} for six different grain sizes, $s = [0.01,0.1,0.4,0.7,0.9,1.2]\,\mu$m, and plot the resulting velocity and density profiles for both gas and dust at $R = 5\,$AU as a function of $z$ in \cref{fig:grain_size}. There is a steady decline in entrainment with increasing grain size. We have selected these sizes in order to have at least one representative sample from each of our earlier defined entrainment subclasses:
\begin{description}[leftmargin=2\parindent,labelindent=2\parindent,itemindent=-2\parindent]
	\item{\makebox[\widthof{\emph{Partial:}}][r]{\emph{Perfect:}}}	The $0.01\,\mu$m grains mirror the gas velocity and density profiles almost exactly.
	\item{\makebox[\widthof{\emph{Partial:}}][r]{\emph{Well:}}}	The $0.1\,\mu$m grains always have a positive acceleration, but noticeably deviate from the gas solution.
	\item{\makebox[\widthof{\emph{Partial:}}][r]{\emph{Weak:}}}	Both the $0.4$ and $0.7\,\mu$m grains exhibit a tell-tale dip (peak) in their velocity (density) profiles.
	\item{\makebox[\widthof{\emph{Partial:}}][r]{\emph{Partial:}}}	The velocity for the $0.9\,\mu$m grains goes negative, causing them to stall above the disc surface. They cannot escape or set up a steady-state outflow.
	\item{\makebox[\widthof{\emph{Partial:}}][r]{\emph{Non:}}}	The initial velocity of the $1.2\,\mu$m grains is negative so that not even partial ejection can be achieved.
\end{description}
Also evident in \cref{fig:grain_size} is the fact that all dust velocities converge to that of the gas for large $z$, regardless of grain size. This is even true -- albeit unphysical -- for partially- and non-entrained dust grains at sufficiently large $z$ (see the dotted lines). Finally, at this radius, $s_\text{max} \approx 0.82\,\mu$m and we have verified that $v_\text{d} = \text{d} v_\text{d} / \text{d} z = 0$ at $z_\text{c}= R/\sqrt{2}$. Thus we can see that all of the classifications and asymptotic behaviours we predicted in \cref{sec:asymptotic_behaviour} are indeed valid. 

The range in vertical outflow velocities in \cref{fig:grain_size} result in a unique angular momentum for each grain size. Thus, by analogy to classical projectile motion from a rotating disc, each grain size will follow a unique trajectory with smaller, faster grains extending higher in the flow than those that are larger and slower. The latter may be sufficiently decoupled from the wind that a combination of gravity and disc flaring can lead to recapture at larger radii. Transport and recapture of dust grains via photoevaporative winds could help explain the observed crystalline fractions at large radii, assuming that radial mixing can transport the crystalline grains to the launch point for the winds \citep[e.g.][]{Owen/Ercolano/Clarke/2011a,Hansen/2014}. We caution, however, that hydrodynamic simulations including settling show a smaller gradation in outflow velocities for entrained dust grains \citepalias{Hutchison/etal/2016}. This implies that the redistribution of grains in the disc via photoevaporative winds is not as effective as our unsettled model suggests.

An overdensity occurs near the disc surface whenever there is a local minimum in the velocity profile, a result of the velocity-density relation in \cref{eq:dust_accretion_rate}. A similar result was found by \citet{Miyake/Suzuki/Inutsuka/2016} in magnetically-driven winds. Beyond $z_\text{c}$ where aerodynamic drag takes over as the dominant force, the larger grains experience a greater acceleration due to their large differential velocities with the gas. As a result, their dust densities drop more rapidly than small grains and a seesaw pattern develops with a common pivot point.\footnote{The common pivot point is actually a coincidence. In general crossings are well localised, but occur at different locations.} Small, perfectly-entrained grains that trace the density profile for the gas represent the smallest (largest) possible density that can be achieved by the dust at small (large) $z$. Thus, assuming $\epsilon$ is constant in the wind is a good, fair, and poor approximation for perfectly-, well-, and weakly-entrained dust grains, respectively.

At the other extreme, the sign reversal in the velocity for grains $s \approx 0.8$--$1\,\mu$m  suggests a pile-up occurs just above the disc's surface (below $1\,\mu$m the initial velocities are negative). The opacity created by the structure in the density profile of the wind will affect the flux of radiation through the disc's atmosphere. The feedback that this will have on photoevaporation is complicated and requires proper radiative transfer calculations. Furthermore, the thermal emission and/or scattered light from the dust grains in the flow can have observational signatures unique to photoevaporating discs, as shown by \citet{Owen/Ercolano/Clarke/2011a}. We leave this for future study as this goes beyond the scope of this paper.


\subsection{Disc radius}
\label{sec:radius}

\begin{figure}
	\centering{\includegraphics[width=\columnwidth]{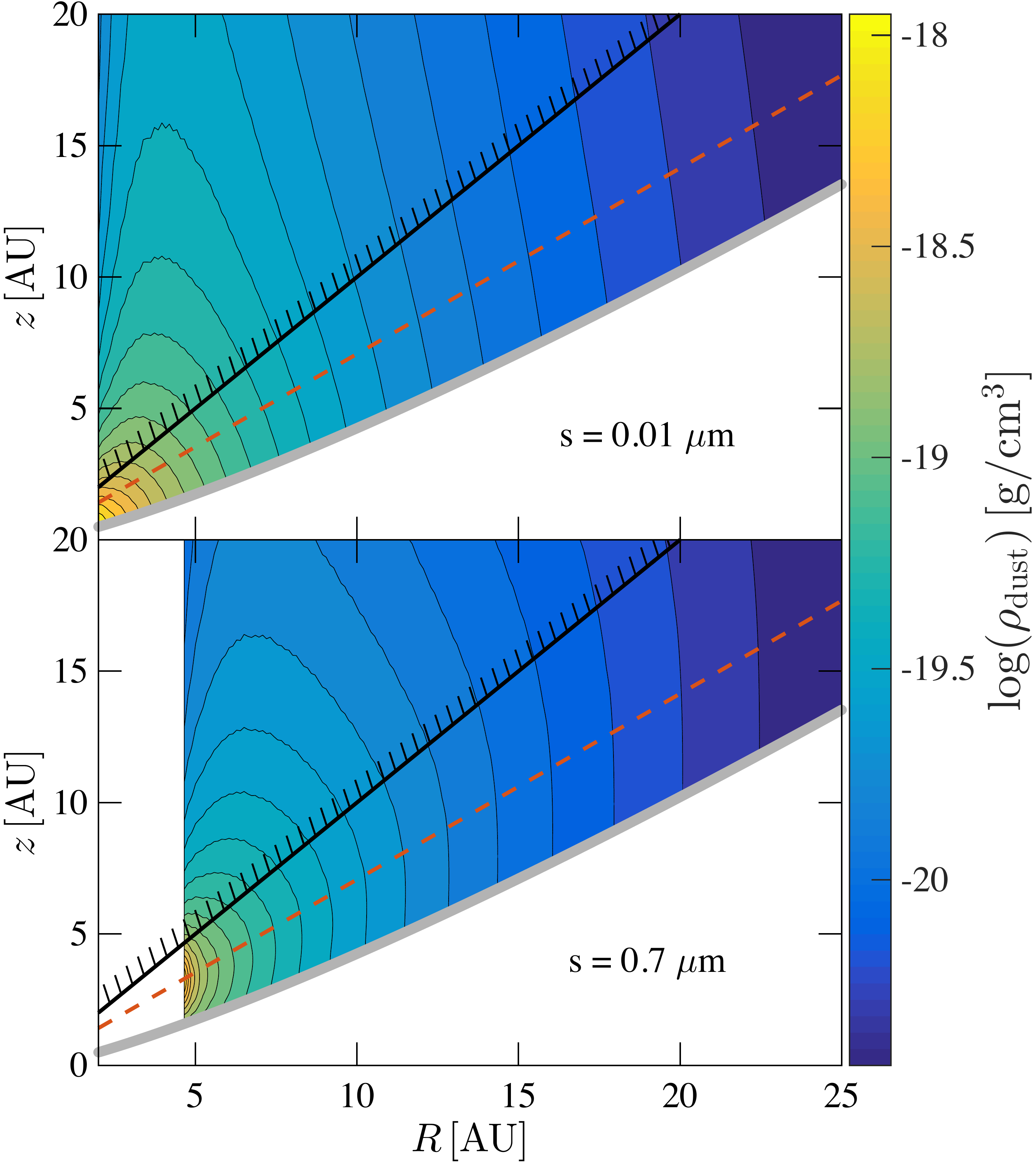}}
	\caption{Density contours for $s = 0.01$ and $0.7\,\mu$m dust grains are created by horizontally stacking 1D calculations at different radii. Empty regions, like the one in the bottom panel at $R \lesssim 5\,$AU, occur whenever $s > s_\text{max}$. When $s \sim s_\text{max}$, the maximum dust density in the wind occurs at $z_\text{c}$ (dashed orange line) rather than along the ionisation front (thick grey line). This suggests that the opacity in the wind may not monotonically decrease with $z$. The solid black line ($z = R$) and hash marks indicate the region where lack of radial pressure gradients and centrifugal motion cause our approximation to break down.}
	\label{fig:radial_solution}
\end{figure}
The assumed underlying 2D disc structure makes our calculations radially consistent at the base of the flow so we can approximate the 2D dust density in the wind by horizontally stacking vertical density maps from different radii. \Cref{fig:radial_solution} compares density contours for two different grain sizes, $s= 0.01$ and $0.7\,\mu$m. However, beyond $z \sim R$ differential pressure and centrifugal effects become significant and our approximation breaks down \citep{Hollenbach/etal/1994}. The hash marks on the black line, $z = R$, indicate the region where our approximation likely breaks down. In contrast to smaller grains, whose density in the wind monotonically declines with $z$, large grains with $s \sim s_\text{max}$ tend to have strongly peaked dust densities with maxima near $z \sim z_\text{c}$.

\begin{figure}
	\centering{\includegraphics[width=\columnwidth]{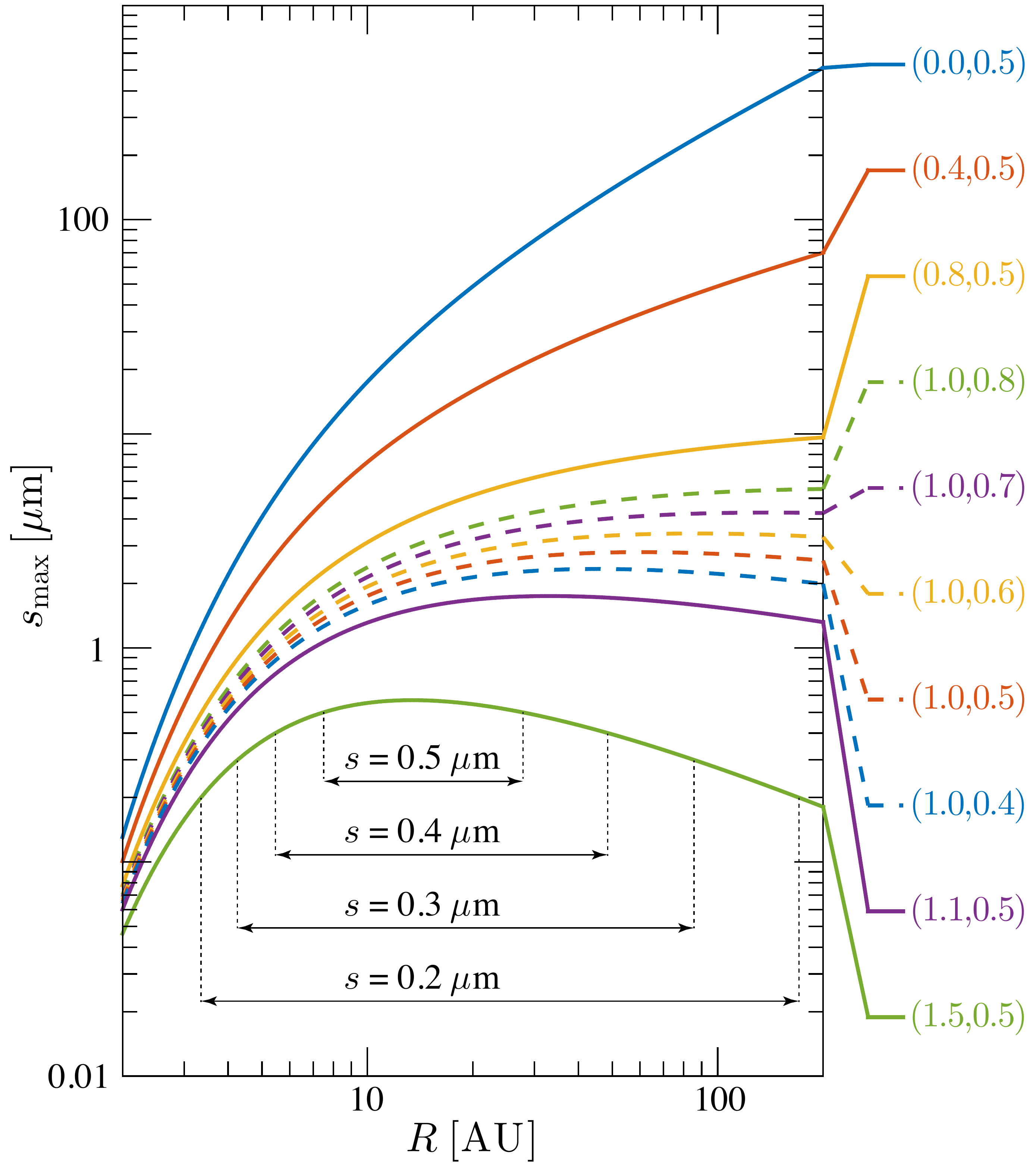}}
	\caption{The maximum entrainable grain size plotted as a function of radius for the different surface density and temperature power law exponents $(p,q)$ labeled next to each curve. Arrows indicate the width of the entrainment region for a few grains in the $(p,q) = (1.5,0.5)$ disc.}
	\label{fig:size_vs_radius}
\end{figure}
\Cref{fig:radial_solution} shows that the entrainment region is not the same for all dust grains. This radial size sorting of dust, first pointed out by \citet{Owen/Ercolano/Clarke/2011a}, is nicely picked up by the analytic expression for the maximum grain size in \cref{eq:smax}. A better illustration of this radial sorting is obtained by plotting $s_\text{max}$ as a function of $R$, as shown in \cref{fig:size_vs_radius} for five disc profiles varying $p \in [0,1.5]$ while holding $q=0.5$ and five disc profiles varying $q \in [0.4,0.8]$ while holding $p=1$. Physically, the exact shape of $s_\text{max}$ is determined by the relative rate of decline between gravity and density as a function of $R$ and $z$. Thus it comes as no surprise that $s_\text{max}$ has a strong dependence on both $p$ and $q$, particularly at large radii where they affect the disc structure the most.

Usually grains with $s < \text{max}(s_\text{max})$ have a unique inner and outer entrainment radius beyond which they cannot be dragged into the flow, but there are a few exceptions. First, because photoevaporation cannot operate below the so-called `critical-radius'  \citep{Liffman/2003,Adams/etal/2004,Alexander/etal/2014},
\begin{equation}
	R_\text{c,EUV} \simeq 1.8 \left( \frac{M}{M_\odot} \right)\,\text{AU},
	\label{eq:r_crit}
\end{equation}
all grains with $s<s_\text{max}$ will share the same inner entrainment radius. A similar situation occurs for the outer entrainment radius if the disc is truncated at the outer edge \citep[e.g. by external photoevaporation; see][]{Facchini/Clarke/Bisbas/2016}. For completeness' sake, we also mention that $s_\text{max}$ has no extremum when $q \geq 2p$, suggesting there is no outer entrainment radius for grains. However, the reality is that discs are finite and $s_\text{max}$ will eventually drop to zero regardless.


\subsection{Base flow density}
\label{sec:base_flow_density}

\begin{figure}
	\centering{\includegraphics[width=\columnwidth]{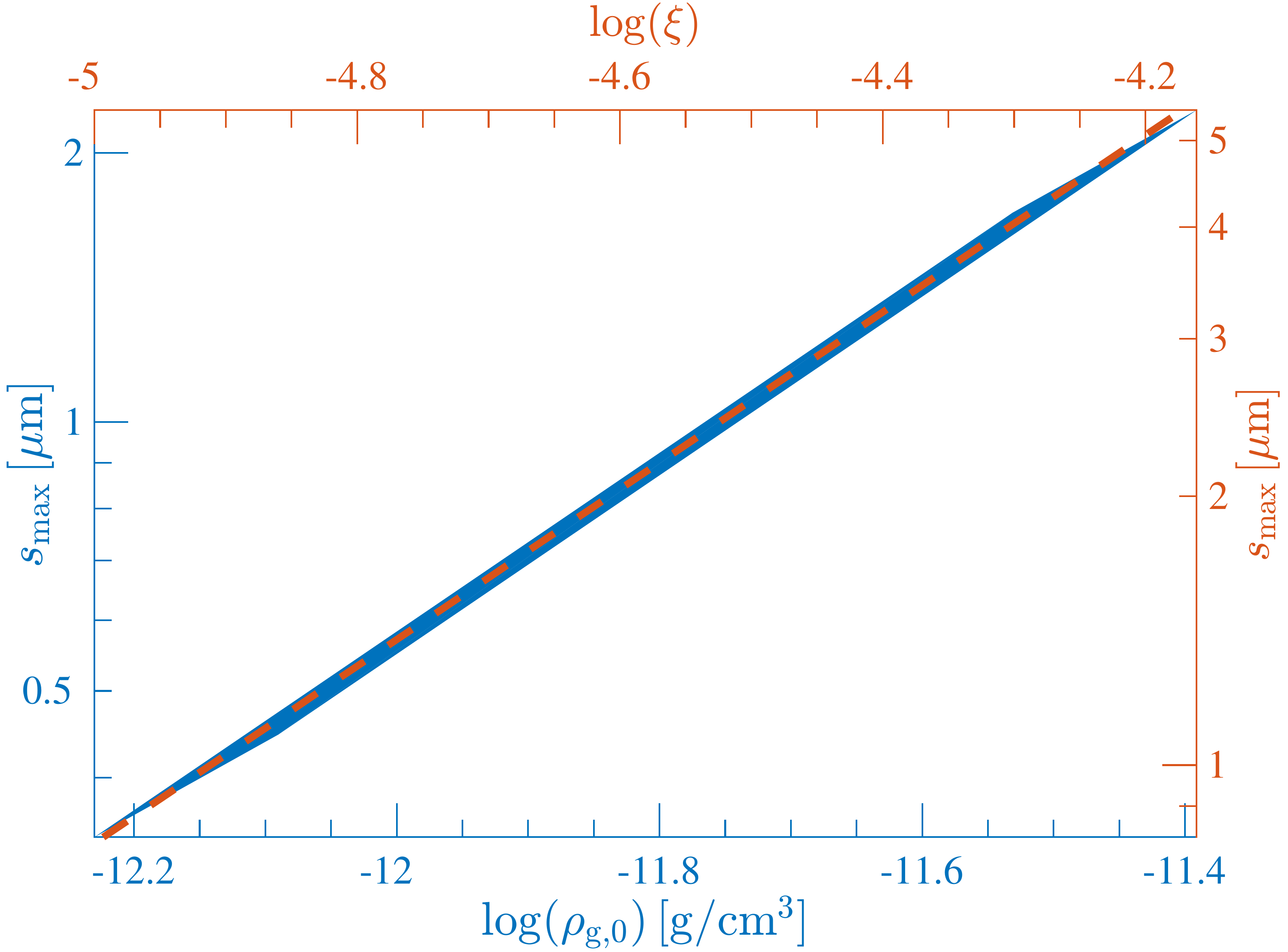}}
	\caption{The maximum entrainable grain size at $R = 5\,$AU as a function of $\rho_\text{g,i}$ (holding $\Sigma_{\text{g},1\text{AU}} = 100\,\text{g/cm}^2$; blue) and $\rho_\text{g,0}$ (holding $\xi = 10^{-5}$; orange). This confirms the linear density-entrainment relationship found by \citetalias{Hutchison/etal/2016} and shows that it is robust against changes to the disc profile.} 
	\label{fig:size_vs_rho}
\end{figure}

The penetration depth has a strong effect on dust entrainment by determining $\rho_\text{g,i}$ and $z_\text{i}$. Rigorously, $\xi$ also influences $v_\text{g,i}$ because the initial velocity is obtained by integrating backwards from the sonic point to $z_\text{i}$. However, the disc density profile is so steep and the velocity profile is so shallow that the actual impact of $\xi$ on $v_\text{g,i}$ is small. Because $\xi$ is directly proportional to the initial base flow density -- and since $\rho_\text{g}$ and $s$ are inversely proportional in the Epstein drag regime -- $\xi$ scales the entrainment properties linearly in $s$, as shown by the orange dashed line and axes in \cref{fig:size_vs_rho}. Note this scaling has no effect on the actual shape of the velocity/density profiles, just a vertical offset in the densities.

Varying $\Sigma_\text{g,1AU}$ has no other effect than to scale $\rho_\text{g,0}$, and hence $\xi$. The exponents $p$ and $q$ have a similar effect. The blue shaded region and axes in \cref{fig:size_vs_rho} illustrate the entire range of midplane densities produced by $p \in [0.5,1.5]$ and $q \in [0.4,0.8]$ at $R=5\,$AU and their associated values of $s_\text{max}$ assuming $\xi = 10^{-5}$. The range of $(p,q)$ pairs gives the data a slight spread, but the average slope of $\rho_\text{g,0}$ vs $s_\text{max}$ is almost identical to that obtained for $\xi$. This is not true for $H_\text{g,1AU}$ and $H$, however. In fact, $H$ has the opposite effect because it is inversely proportional to $\rho_\text{g,0}$. Smearing the same density over a larger volume results in a diminished base flow density and a smaller maximum grain size $s_\text{max}$.

\subsection{Stellar mass}
\label{sec:stellar_mass}

\begin{figure}
	\centering{\includegraphics[width=\columnwidth]{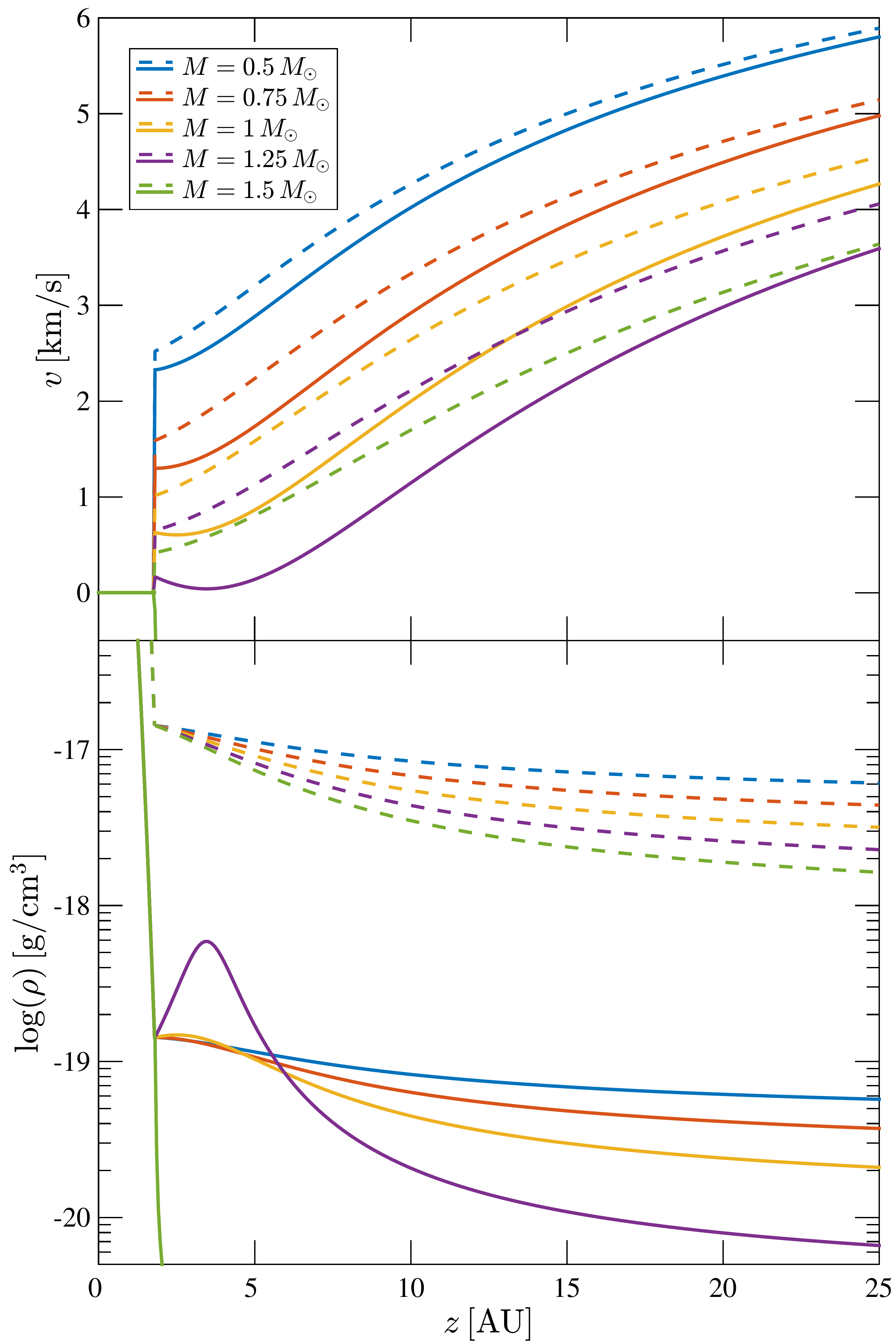}}
	\caption{Similar to \cref{fig:grain_size}, but varying stellar mass instead of grain size. Gas and dust properties are represented using dashed and solid lines, respectively. The grain size for all dust curves is $s=0.4\,\mu$m. Increasing the stellar mass reduces the initial outflow velocities of both gas and dust. As a result, fewer grain sizes can be entrained in the flow.}
	\label{fig:star_mass}
\end{figure}

\begin{figure}
	\centering{\includegraphics[width=\columnwidth]{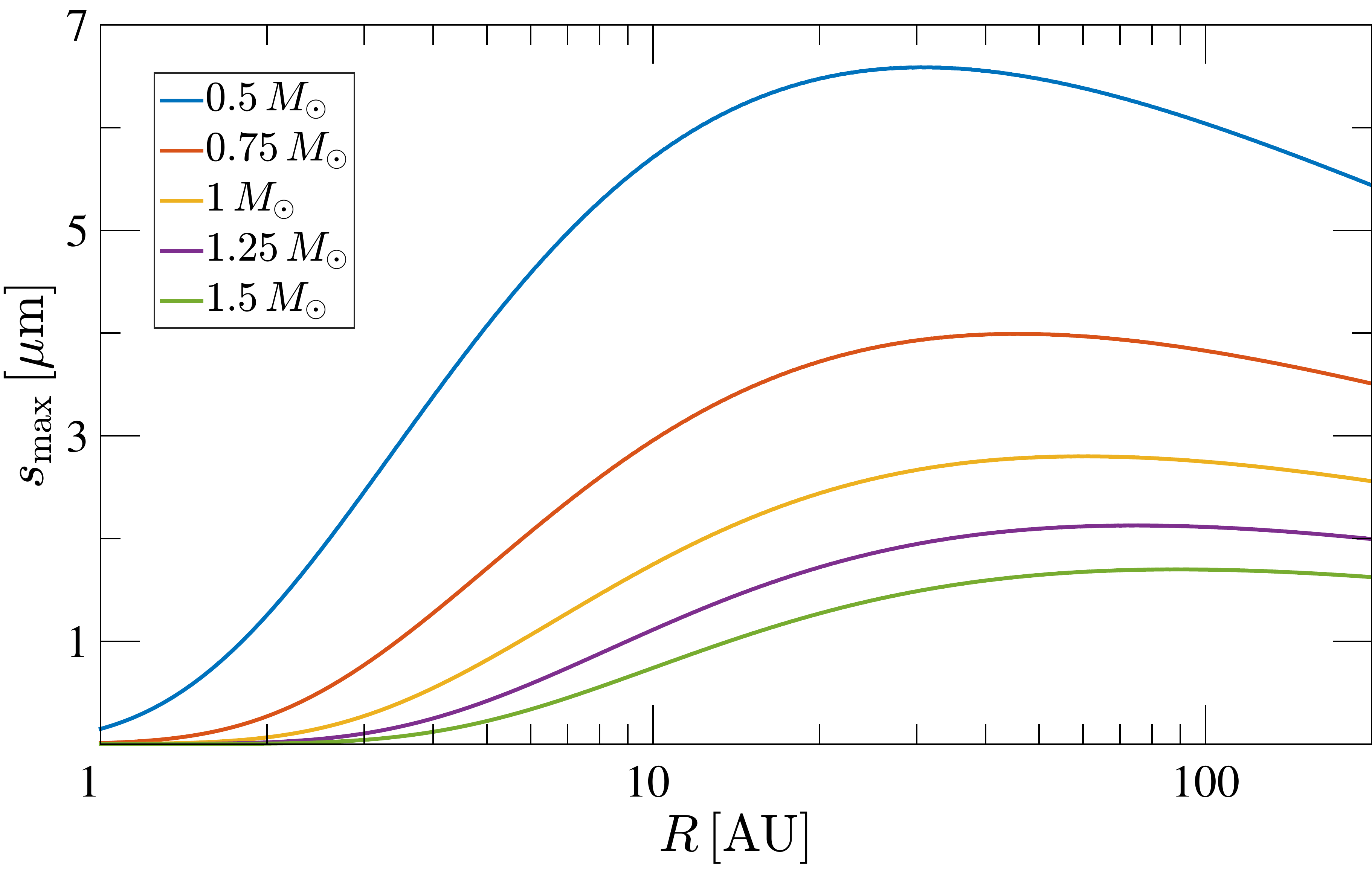}}
	\caption{The maximum entrainable grain size plotted as a function of radius for the stellar masses listed in the plot. The larger gravitational well produced by higher mass stars require more energy to escape, thus prohibiting the larger, slower moving grains from becoming entrained in the outflow.}
	\label{fig:size_vs_mass}
\end{figure}

Massive stars tend to have more massive discs and higher luminosities \citep{Andrews/etal/2013,Andrews/2015}, thereby affecting $\Sigma_\text{g}$ and $\xi$, respectively. However, before addressing these complexities, it is instructive to look at how gravity alone affects dust entrainment in winds. From \cref{eq:accretion_rate} we can see that the overall velocity profile for the gas decreases as stellar mass increases. This is illustrated by the dashed lines in the top panel of \cref{fig:star_mass} using the following stellar masses, $M=[0.5,0.75,1,1.25,1.5]\,M_\odot$. The resulting dust velocities for $s=0.4\,\mu$m (solid lines in top panel) are shown in each case and go from being well-entrained at $0.5\,M_\odot$ to being completely non-entrained at $1.5\,M_\odot$. Furthermore, in \cref{fig:size_vs_mass} we find that $s_\text{max}$ globally decreases as the stellar mass increases. Thus, dust entrainment is inversely proportional to the stellar mass.

The gas and dust densities in the bottom panel of \cref{fig:star_mass} are phenomenologically similar to those in \cref{fig:grain_size}, but illustrate two new points of interest. First, the density gradient in the wind gets steeper as the mass of the star increases. This suggests that the gas/dust envelope surrounding photoevaporating discs is smaller for host stars that are more massive. Second, placing identical dust distributions around stars of different masses will produce different density profiles in the wind. The weakened entrainment at higher stellar masses will result in diminished outflow velocities and higher wind densities. Therefore, although we expect a narrower grain size distribution in the wind, their densities will be enhanced compared to the same grains at lower stellar mass.

We can now see that increasing the stellar mass weakens dust entrainment while increasing disc mass and/or stellar luminosity strengthens it. It is possible to tune these three parameters to minimise their combined effect on entrainment, but there is always a residule density signature. For example, compensating for the weakened entrainment around a high mass star by increasing $\rho_\text{g,0}$ and/or $\xi$ would, according to \cref{fig:size_vs_rho}, increase the already inflated wind density. Thus, even in the case where dust entrainment is minimally affected, higher mass stars should exhibit higher dust densities in their outflows. 


\section{Effects of dust settling}
\label{sec:effects_of_settling}

In this section we use a turbulent disc model to derive the dust density for a settled disc. With this density relation, we derive a new constraint on the maximum grain size in winds that reflects how some entrainable dust grains settle well below the launch point for the flow. Finally, we discuss how settling affects the results in \cref{sec:results}.

\subsection{Turbulent disc model}
\label{sec:turbulent_disc_model}

\begin{figure}
	\centering{\includegraphics[width=\columnwidth]{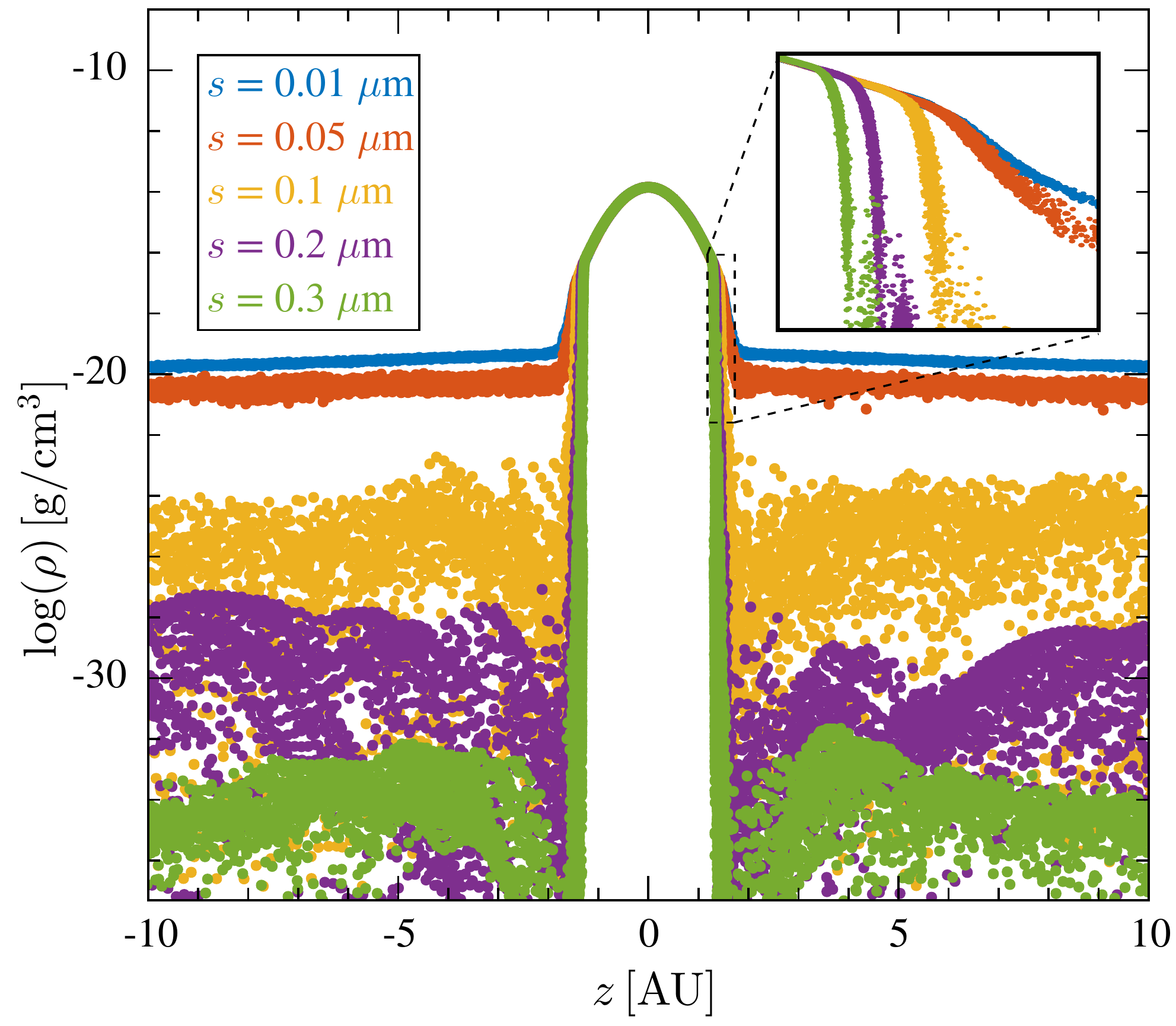}}
	\caption{Dust densities for the entrained dust grains in the hydrodynamic simulations run in \cref{sec:verifying_smax}. The semi-analytic model can predicted the shape of the density profile because $\rho_\text{d} \propto 1/v_\text{d}$, but it cannot account for the initial density offset for each grain size in a settled disc without first knowing the density of each grain size at the ionisation front.}
	\label{fig:density_with_inset}
\end{figure}

We saw in \cref{sec:verifying_smax}, that settling does not affect the accuracy of the semi-analytic model in determining $s_\text{max}$ or the velocity of entrained dust grains -- as long as there is a non-zero dust density at the ionisation front. With the velocities pinned down, the continuity equation ensures that the shape of the density profile is known too. However, \cref{fig:density_with_inset} shows that dust settling within the disc will result in a size dependent $\rho_\text{d,i}$ that requires external calibration. In principle, this is similar to the $40\%$ shift we applied earlier to $\rho_\text{g,i}$, but is made complicated by the unique interaction each grain size has with the disc. Calibrating $\rho_\text{d,i}$ from simulations would render the model superfluous, so we approximate the density using a turbulent disc model.

The continuity approximation used throughout \cref{sec:results} can already be thought of as extreme turbulence in the disc. Here we just switch to a more physical description by assuming a finite turbulent viscosity, $\nu_\text{t} = \alpha c_\text{s}^2/\Omega_\text{K}$ where $\alpha$ is a dimensionless constant \citep{Shakura/Sunyaev/1973}. Following \citet{Dubrulle/Morfill/Sterzik/1995}, the vertical distribution of dust density in a turbulent disc governed by
\begin{equation}
	\frac{\partial \rho_\text{d}}{\partial t} + \nabla(\rho_\text{d} \Delta \mathbf{v}) = \nabla \left[ \rho_\text{g} \kappa_\text{t} \nabla \left( \frac{\rho_\text{d}}{\rho_\text{g}}\right)  \right],
	\label{eq:full_turb_evolution}
\end{equation}
with the turbulent diffusivity, $\kappa_\text{t}$, is given by 
\begin{equation}
	\kappa_\text{t} = \frac{\Omega_\text{K} \alpha H^2}{\sqrt{1+\Gamma}},
	\label{eq:small_kappa_t}
\end{equation}
where $\Gamma = 5/3$ for isotropic, incompressible turbulence. \Cref{eq:small_kappa_t} assumes that $\Omega_\text{K} t_\text{s} \ll 1$, which is true for all grain sizes considered in this study. Noting that (i) the vertical timescale is much shorter than the radial timescale; (ii) assuming that the dust is always small enough to settle at the local terminal velocity (i.e. $\Delta v_z = -z \Omega_\text{K}^2 t_\text{s}$); and (iii) restricting ourselves to stationary solutions allows us to rewrite \cref{eq:full_turb_evolution} as a separable first-order differential equation,
\begin{equation}
	-z \Omega_\text{K}^2 t_\text{s} \rho_\text{d} = \kappa_\text{t} \left(  \frac{z}{H^2} \rho_\text{d} + \frac{\text{d} \rho_\text{d}}{\text{d} z}  \right),
	\label{eq:simp_turb_evolution}
\end{equation}
which has the solution
\begin{equation}
	\rho_\text{d} = \varepsilon_0 \rho_\text{g} \exp \left[-\frac{\sqrt{1+\Gamma}}{\alpha} \frac{\rho_\text{grain} s \Omega_\text{K} }{ \rho_\text{g,0} c_\text{s}} \left(\frac{\rho_\text{g,0}}{\rho_\text{g}}-1\right) \right].
	\label{eq:small_kappa_dust_dens_sol}
\end{equation}

\subsection{Settling-limited maximum grain size}
\label{sec:the_settling-limited_maximum_grain_size}

Our hydrodynamic simulations are non-turbulent, but they can provide an order of magnitude estimate of the $\alpha$ needed in this model to produce a realistic stratified dusty disc. Using $\alpha = 0.05$, \cref{eq:small_kappa_dust_dens_sol} reproduces a density structure that is similar to that shown in \cref{fig:density_with_inset}. Although the dust density in the model and simulation can drop to arbitrarily low values, it is natural to believe that below some threshold the density becomes physically irrelevant. Alternatively, the threshold could be caused by a technological limitation in the observing power of a particular telescope. Either way, limiting $\rho_\text{d}$ allows us to invert \cref{eq:small_kappa_dust_dens_sol} and solve for the maximum grain size, $s_\text{tur}$, allowed/observed to enter the wind due to settling in the disc:
\begin{equation}
	s_\text{tur} = \frac{\alpha}{\sqrt{2 \pi (1+\Gamma)}} \left( \frac{ \xi}{1-\xi}  \right) \left( \frac{ \Sigma_\text{g}}{\rho_\text{grain}} \right) \log \left(\frac{\varepsilon_0 }{\varepsilon_\text{i}}\right),
	\label{eq:s_tur}
\end{equation}
where we have defined $\varepsilon_\text{i} \equiv \rho_\text{d}(z_\text{i})/ \rho_\text{g}(z_\text{i})$.

In this study, we are more interested in the fundamental process by which settling limits the grains in the flow than by observational feasibility; therefore, we define the above threshold as the point where $100 \times (1-\beta)\%$ of the grains are contained within $|z| < z_\text{i}$. For this to be meaningful, we assume $0 < \beta \ll 1$. The role of $\beta$ in this analysis is to determine the likelihood that a given grain size will have settled below the ionisation front and not be entrained in the wind. Mathematically, we do this in three steps: (i) we normalise the density to get the probability density function (PDF), (ii) we integrate the PDF over the interval $[-z_\text{i},z_\text{i}]$ and set it equal to the desired threshold fraction of grains in the disc:
\begin{equation}
	\frac{\int_{-z_\text{i}}^{z_\text{i}} \rho_\text{d}(z,s) \, \text{d}z}{\int_{-\infty}^{\infty} \rho_\text{d}(z,s) \,\text{d}z} = 1-\beta,
	\label{eq:cdf_function}
\end{equation}
and (iii) we solve for the critical value $s$ that makes this relation true. The family of solutions for \cref{eq:cdf_function} is indeed given by \cref{eq:s_tur}, but $\varepsilon_\text{i}$ is a non-linear function of both $\xi$ and $\beta$. In the interest of keeping $s_\text{tur}$ analytic, we fit $\varepsilon_\text{i}$ with a polynomial surface
\begin{equation}
	\varepsilon_\text{i} \simeq \sum_{m=0}^2 \sum_{n=0}^3 C_{mn} \xi^m \beta^n,
	\label{eq:varepsilon}
\end{equation}
with coefficients $C_{mn}$ given in \cref{tab:coefficients}. The root-mean-square error for this fit is $\sim1\%$.
\begin{table}
	\centering
	\caption{Coefficients $C_{mn}$ for the fit of $\epsilon_\text{i}$ in \cref{eq:varepsilon}.}
	\label{tab:coefficients}
\sisetup{output-decimal-marker = {.}}
\begin{tabular*}{\columnwidth}{@{\extracolsep{\stretch{1}}} c  S[table-format=1.5] S[table-format=1.5] S[table-format=1.5] S[table-format=1.6] @{}} \toprule
		\diag{.01em}{.3cm}{$m$}{$n$}	&	{0}					&	{1}				&	{2}					&	{3}  			\\ \midrule
		{0}						&	-1.04600				&	0.85290			&	-0.02126				&	-0.001092		\\
		{1}						&	-0.80850				&	 0.08992			&	0.00497				&	0			\\
		{2}						&	-0.06833				&	 -0.00408			&	0					&	0			\\ \bottomrule
\end{tabular*}
\end{table}

\begin{figure*}
	\centering{\includegraphics[width=\textwidth]{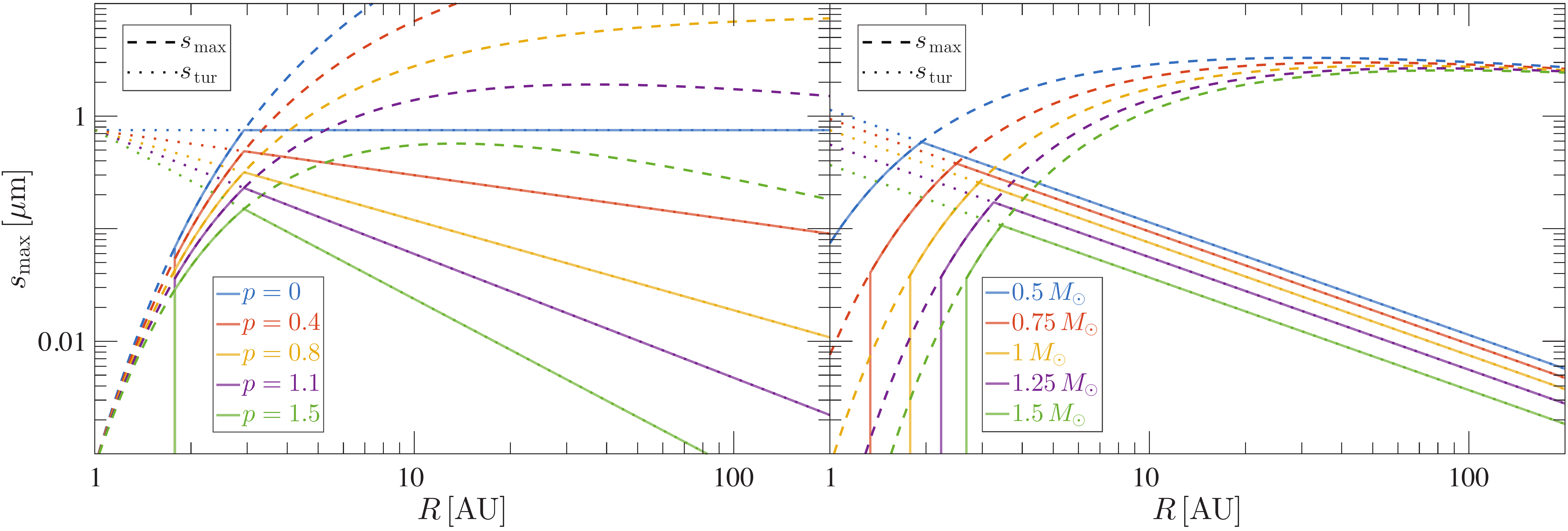}}
	\caption{The maximum grain size in the wind as a function of radius while varying $p$ (left) and $M$ (right) assuming a minimum dust-to-gas ratio, $\varepsilon_\text{i} = 10^{-10}$. The dotted and dot-dashed lines show the maximum entrainable grain size with and without settling, respectively. The true value of $s_\text{max}$ is the minimum of these two curves. In the right panel we assume that both $\xi$ and $\Sigma_\text{g,1AU}$ have a $\sqrt{M/M_\odot}$ dependence. The stark cutoff in the inner disc is due to quenching of photoevaporation inside of $R_\text{c,EUV}$ in \cref{eq:r_crit}. In every case, settling severely limits the value of $s_\text{max}$ in the wind.}
	\label{fig:combined_settled_smax_relation}
\end{figure*}

To keep our analysis as general as possible and still be able to make informed predictions about the maximum grain size found in winds, we restrict ourselves to using only the highest confidence levels. The more strict the upper limit we set on $s_\text{tur}$ in this model, the more likely it will apply in other turbulent disc models, albeit as a softer limit. Numerically, the smallest value of $\beta$ we can use before experiencing roundoff errors is $10^{-15}$. This corresponds to a $\gtrsim 6\sigma$ confidence level that $s_\text{tur}$ has settled below the ionisation front. The analytic form of $s_\text{tur}$ allows us to decrease $\beta$ much lower, revealing that $s_\text{tur}$ asymptotically reaches a maximum value. However, extrapolating far beyond the fitted data often leads to erroneous results, so we feel it is sufficient to assume $\beta = 10^{-15}$.

Although we believe these predictions are representative of discs undergoing photoevaporation, the exact values and likelihoods are biased by the assumptions of our turbulent disc model. Turbulence in the upper disc is not well understood and could be different to what we assume here. Furthermore, recent non-ideal magneto-hydrodynamic simulations suggest that the accretion stress may be largely laminar in the inner $\lesssim 30\,$AU of the disc \citep[e.g.][]{Bai/2013,Bai/2014,Lesur/Kunz/Fromang/2014,Gressel/etal/2015,Simon/etal/2015}, thereby reducing values of $\alpha$ controlling turbulent mixing in these regions. This would result in even smaller values of $s_\text{tur}$, i.e. a smaller reservoir of grains to be carried out by the wind.. Finally, our analysis does not take into account the collisional evolution and redistribution of small dust grains, which can potentially trap small grains in a layer significantly thinner than the gas \citep{Krijt/Ciesla/2016}. This could similarly reduce $s_\text{tur}$ below that implied by \cref{eq:s_tur}.

\subsection{Modification to earlier results}
\label{sec:modification_to_earlier_results}

Given this new constraint, the true maximum grain size in the wind is given by the minimum of $s_\text{max}$ and $s_\text{tur}$. \Cref{fig:combined_settled_smax_relation} shows the effect this new relation has on our earlier calculations. The left panel varies the disc's surface density exponent $p$ while the right panel varies stellar mass (note $s_\text{tur}$ does not depend on $q$). The dashed and dotted lines show $s_\text{max}$ and $s_\text{tur}$ respectively, while the solid lines indicate the true maximum grain size entrained at each radius. In the right panel we have assumed that both $\xi$ and $\Sigma_\text{g,1AU}$ are multiplied by a mass factor, $\sqrt{M/M_\odot}$. The mass factor here is simply illustrative; without a mass dependence, the curves would be colinear. As mentioned earlier, any positive correlation like this between stellar mass and $\xi$ and/or $\Sigma_\text{g,1AU}$ will increase the dust density in the wind. Thus, despite both $s_\text{max}$ and $s_\text{tur}$ being inversely proportional to stellar mass, when taking into account that telescopes have a limited resolution, the higher wind densities at higher stellar masses may result in a larger range of observable grain sizes in the wind (as opposed to entrainable grain sizes). 

One of the most striking features of \cref{fig:combined_settled_smax_relation} is that dust settling is the dominant mechanism determining the maximum grain size in the wind at almost all radii in the disc. Moreover, since photoevaporation only operates at $R > R_\text{c,EUV}$, the region dominated by $s_\text{max}$ can be very small. This is an important result because it means that only dust in the inner few AU of the disc will experience weak entrainment in the winds. Consequently, processes reliant on slow moving dust grains, such as mass transport to the outer disc and pile-ups at the disc surface, are restricted to the inner disc. Note, this does not rule out photoevaporation as the transport mechanism for the earlier mentioned crystalline grains since these grains are formed in situ in the inner disc.

The values used for $\xi$ and $\Sigma_\text{g,1AU}$ are fairly conservative and increasing either one will lead to larger grains entrained in the outflow. However, since both $s_\text{max}$ and $s_\text{tur}$ are affected similarly, the transition from wind-limited to settling-limited flow remains largely unaffected. The slight change that does occur is because the dust's density profile is no longer a perfect gaussian like that of the gas. Ultimately, this distorts the linear density-entrainment relationship discussed earlier, but the deviation is small and decreases as $\xi$ increases.


\section{Implications for X-ray and FUV photoevaporation}
\label{sec:implications}

Strictly speaking, the results above are only applicable to EUV photoevaporation, but the flows produced by EUV, X-ray, and FUV radiation are all hydrodynamically similar. Therefore, we expect dust grains entrained by X-ray or FUV driven winds to behave qualitatively like the dust grains described in this paper. However, there are some important differences to consider that can influence the way dust grains behave in winds \citep[see][]{Alexander/etal/2014,Gorti/etal/2016}. EUV radiation accelerates winds faster than the other energy regimes because it almost instantaneously heats winds to $\sim 10^4\,$K. Heating due to X-rays and FUV radiation is more gradual and cooler. Maximum X-ray temperatures reach $\sim 3000$--$5000\,$K while FUV temperatures range from $>1000\,$K in the inner disc to $\lesssim100\,$K in the outer disc. In contrast, the energy regimes rank in opposite order in terms of penetration as evidenced by their associated penetration columns of neutral hydrogen: $N_\text{H,EUV} \sim 10^{18}$--$10^{20} \, \text{cm}^{-2}$, $N_\text{H,X-ray} \sim 10^{21}$--$10^{22} \, \text{cm}^{-2}$, and $N_\text{H,FUV} \sim 10^{21}$--$10^{23} \, \text{cm}^{-2}$ \citep{Ercolano/Clarke/Drake/2009,Alexander/etal/2014}).

While the lower outflow velocities (due to lower temperatures) achieved by X-ray and FUV radiation reduce entrainment efficiency by at least a factor of a few \citepalias{Hutchison/etal/2016}, according to \cref{sec:base_flow_density} the significantly larger penetration depths likely overcompensate and produce better entrainment overall, raising the value of $s_\text{max}$. At the same time, by penetrating deeper into the disc, X-ray and FUV winds have access to larger, more settled grains than EUV winds. As a result, $s_\text{tur}$ also increases and we may expect the same qualitative behaviour as shown in \cref{fig:combined_settled_smax_relation}, but shifted to larger grain sizes.

The indirect thermal coupling with the dust should be minimal for X-rays since they are almost unaffected by small dust grains in the upper atmospheres of discs \citep[see][]{Gorti/etal/2016}. However, X-rays are primarily absorbed by heavy elements in the disc -- many of which can be found in dust grains -- and the effects of dust settling on X-ray driven photoevaporation have not been studied in detail \citep{Alexander/etal/2014}. FUV winds are much more likely to be affected by changes in the dust phase. \citet{Gorti/Hollenbach/Dullemond/2015} report noticeable reductions in disc lifetimes when dust evolution (i.e. settling, migration, and coagulation/fragmentation) is considered alongside FUV photoevaporation. While increased mass-loss rates imply better entrainment, the radial variation in FUV wind temperatures and fragmentation efficiency (i.e. replenishment of small grains), plus the tendency for dust to migrate inwards, suggests the increase does not occur uniformly across all radii. Furthermore, the opacity from entrained dust grains has never been modeled self-consistently and could impede FUV photoevaporation rates, especially in the outer disc. Further studies need to be conducted in order to determine the relative importance of each of these processes and how they affect the radial profile of grain sizes carried into the winds.


\section{Conclusions}
\label{sec:conclusions}

We have developed a simple but powerful model using a non-rotating, plane-parallel, photoevaporating atmosphere to estimate the conditions by which dust grains can be carried into a photoevaporative wind. \Cref{eq:dust_ode} gives the maximum grain size, $s_\text{max}$, that can be entrained by the outflow. The model accurately recovers almost all of the results from our more rigorous hydrodynamic simulations in \citetalias{Hutchison/etal/2016} for different stellar and disc parameters. This implies that any observational constraint on the grain size in the wind can be translated into a strict lower bound on the mass loss rate of the disc, \cref{eq:surface_density_mass_loss}.

This relation is consistent with, but more versatile than earlier estimates of $s_\text{max}$ by \citet{Takeuchi/Clarke/Lin/2005} and \citet{Owen/Ercolano/Clarke/2011a}. We show in particular that $s_\text{max}$ varies with disc radius with $\text{max}(s_\text{max}) < 10 \,\mu$m for typical T Tauri stars. However, the largest grain size entrained in the flow may be much smaller than $s_\text{max}$, since dust settling prevents the replenishment of large grains in the wind (except in the inner few AU of the disc). In addition to determining the maximum entrained grain size, we uncover five distinct behavioural classes of dust grains in photoevaporative winds: perfectly-, well-, weakly-, partially-, and non-entrained grains. These five classes exhibit different outflow velocities, which may lead to stratified dust layers in the wind and recapture of weakly-entrained dust grains by the disc at large radii.

The stellar mass has a non-linear relationship with $s_\text{max}$ that alters the shape of the velocity and density profiles for the gas and dust. We find that more massive stars will tend to host winds with a more compact envelope, higher wind density, and higher dust-to-gas ratio. Although the maximum entrainable grain size decreases with increasing stellar mass, any positive correlation between luminosity and/or disc mass with the mass of the central star may result in a larger range of observable grain sizes at higher stellar mass.

\section*{Acknowledgements} 
We thank Daniel Price for useful discussions and the anonymous referee whose comments helped improve this paper. M.H. acknowledges funding from a Swinburne University Postgraduate Research Award (SUPRA) and support from the Astronomical Society of Australia. G.L. acknowledges funding from the European Research Council for the FP7 ERC advanced grant project ECOGAL. S.T.M. acknowledges partial support from PALSE (Programme Avenir Lyon Saint-Etienne). Simulations were performed on the swinSTAR supercomputer at Swinburne University of Technology.




\bibliographystyle{mnras}
\bibliography{$HOME/Dropbox/Bibtex_library/library}





\bsp	
\label{lastpage}
\end{document}